\documentclass[12pt,a4paper,dvips]{scrartcl}
\usepackage[ansinew,latin1]{inputenc}
\usepackage[T1]{fontenc}
\usepackage{graphicx}
\usepackage{amsmath}
\usepackage{amssymb}

\usepackage{latexsym}
\usepackage{feynmf}     
\usepackage{sparticles}

\begin{document}

\begin{titlepage}
\begin{center} {\LARGE \bf Low-energy sector of 8-dimensional General Relativity: Electro-Weak model and neutrino mass 
\vspace{0.2in} \\}\vspace*{1cm}
{\bf Francesco Cianfrani$^1$, Giovanni Montani$^{123}$}\\
\vspace*{1cm}
$^{1}$ICRA---International Center for Relativistic Astrophysics\\
Dipartimento di Fisica (G9), Universit\`a  di Roma, ``Sapienza",\\
Piazzale Aldo Moro 5, 00185 Rome, Italy.

\vspace*{0.5cm}

$^{2}$ENEA C.R. Frascati (Dipartimento F.P.N.),
Via Enrico Fermi 45, 00044 Frascati, Rome, Italy.

\vspace*{0.5cm}

$^{3}$ICRANet C. C. Pescara, Piazzale della Repubblica, 10, 65100 Pescara, Italy.

\vspace*{0.5cm}

e-mail: montani@icra.it\\
        francesco.cianfrani@icra.it\\
\vspace*{1.8cm}

 \vspace*{1cm} \\

\vspace*{1cm}

{\bf  Abstract\\} \end{center}
In a Kaluza-Klein space-time $V^{4}\otimes S^{1}\otimes S^3$, we demonstrate that the dimensional reduction of spinors provides a 4-field, whose associated $SU(2)$ gauge connections are geometrized. However, additional and gauge-violating terms arise, but they are highly suppressed by a factor $\beta$, which fixes the amount of the spinor dependence on extra-coordinates. The application of this framework to the Electro-Weak model is performed, thus giving a lower bound for $\beta$ from the request of the electric charge conservation. Moreover, we emphasize that also the Higgs sector can be reproduced, but  neutrino masses are predicted and the fine-tuning on the Higgs parameters can be explained, too.

\end{titlepage}

\section{Introduction}
One of the main issues of modern Physics is to recast all interactions into a unified picture. This aim requires to fix a new paradigm of our knowledge, able to explain the deep difference between gravity, geometrized by General Relativity, and strong-electro-weak interactions, having the form of gauge theories.\\
We have two different ways of accomplishing the above purpose: to recover even gravity as a gauge theory (an example is Poincar\'e Gauge Theory \cite{1}) or to search for a geometrical formulation of the Standard Model. Since a gauge theory for gravity seems to work only in a quantum level \cite{Pol}\cite{Rov}, then  a unification picture based on a geometrical point of view is suitable for a classical formulation.\\
In this paper, we consider a Kaluza-Klein (KK) approach for the geometrization of the Electro-Weak model. This scheme \cite{2} \cite{3} (for a review see \cite{4}, \cite{5} or \cite{6}) pursues the Einstein's idea to give a physical content to all the metric degrees of freedom; in particular, it is based on recognizing our four-dimensional space-time as embedded in a multidimensional one, whose additional (off-diagonal) metric components determine gauge bosons. To reconcile this scheme with our four-dimensional phenomenology, the extra-space is assumed to be compactified at distances forbidden to experiments. Therefore, our all-day Physics is a low-energy approximation with respect to compactification energy scales.\\ 
This framework was proved to be very useful for bosons, since Yang-Mills Lagrangian comes out from the dimensional reduction of the Einstein-Hilbert action. However the introduction of fields able to reproduce (gauge coupled) four-dimensional fermions is a much more difficult task, except for the Abelian case. In fact, by expanding \textit{al\'a} Fourier functions along the fifth dimension, $U(1)$ transformations can simply be reproduced by $x^{5}$-translations \cite{6}.\\ 
With the aim of dealing with the Electro-Weak model, here we propose a phenomenological approach for a $SU(2)$ gauge interaction: spinors are introduced with an extra-coordinates dependence which is induced by their low-energy character. In this respect, we require that spinors satisfy the Dirac equation when averaged on the extra-space $S^{3}$, thus taking into account the un-observability of this space. Then, we consider an expansion in an order parameter $\beta^{-1}$, characterizing the dependence on extra-coordinates. The geometrization of $SU(2)$ gauge connections is easily accomplished this way at the lowest order; moreover, the first additional terms, of order $\beta^{-1}$, imply violations of gauge symmetries. Therefore, an estimate of the parameter $\beta$, once this framework is applied to the Electro-Weak model and compared with experimental limits, is obtained.\\
Furthermore, we introduce the multidimensional analogous of the Higgs field, by which we can give masses to all particles, including neutrinos; then, its extra-dependence gives a $\phi^{\dag}\phi$ term having a coefficient of the compactification length order. This term can account for the fine tuning required to stabilize the Higgs mass \cite{sus}. Reproducing the spontaneous symmetry breaking mechanism, we also obtain a non-vanishing photon mass, from which a lower bound on $\beta$ is obtained.\\
In particular, the organization of all this topics in the paper is the following: in section 2 we analyze the standard way spinors are introduced in a KK framework; in section 3 we turn to our phenomenological approach: we determine, at the lowest order in $\beta^{-1}$, a solution of the Dirac equation on $S^{3}$, for which in section 4 the geometrization of $SU(2)$ gauge connections is performed; in section 5 we consider the application of previous results to the geometrization of the Electro-Weak model, where the $U(1)$ hypercharge symmetry has to be included; in section 6 non-standard couplings, coming from $\beta^{-1}$ terms, are analyzed;  in section 7 we reproduce the Higgs mechanism, with the new feature of a neutrino mass and a constraint on $\beta$ due to the prediction of a massive photon too; finally, in section 8 brief concluding remarks follow.

\section{Spinors within the Kaluza-Klein framework}
The development of the original KK theory \cite{2} \cite{3} provided a framework where the geometrization of the electro-magnetic field could be performed. After interactions other than gravity were recognized as the Yang-Mills ones, increasing interest in such models appeared. In fact, enlarging the dimensions of space-time via a compact homogeneous manifold, the boson sector of gauge theories can be recovered from the Einstein-Hilbert action \cite{4}, \cite{5}.\\
Problems arise when fermions are introduced, because they are treated as matter fields and relevant shortcomings of the model take place when the following point of view is addressed. In a straightforward approach, multidimensional spinors are provided by simply extending the four-dimensional formalism; this line of thinking has to face non trivial questions, such as the fermion mass, the chirality problem and so on.\\
In fact, eigenvalues of the Dirac equation on the extra-space behave, under the KK hypothesis, like masses and they are of the compactification scale order; this fact leads to the model inconsistency because, for the multidimensional spinors, no zero-eigenvalue state exists \cite{13}. There are two main ways to avoid this result: the introduction of torsion \cite{des} or of non-geometrical gauge fields \cite{bl}. We stress that these possibilities look rather ad-hoc and the last one conflicts with the spirit of KK models, since it introduces non-geometrical bosons.\\
As far as the chirality is concerned, when the Electro-Weak model is considered in KK, different transformation properties for left-handed and right-handed spinors have to be implemented in a geometrical way. Therefore, the hope was that the right-handed and left-handed zero modes of the Dirac operator behaved differently under n-bein rotations. This possibility is ruled out by the Atiyah-Hirzebruch theorem \cite{ah}, despite the case non-geometrical gauge fields are present. This way, the KK program can be applied to spinors at the price of introducing external gauge bosons \cite{bl2}, thus, according to this point of view, the most important result (the geometrization of the boson component) would be destroyed.\\
However, as said above, these results are based on the assumption that multidimensional and four-dimensional spinors coincide. Here we propose a different approach, where 4-spinors are the relic of the full field, after the dimensional reduction has been preformed. In this sense, we develop a phenomenological model, able to reproduce the Electro-Weak theory from standard KK hypothesis.

\section{Dirac equation on the 3-sphere}
Since in a KK framework interactions other than gravity are geometrized (by virtue of off-diagonal metric components), the properties of matter fields are fixed by their dependence on extra-coordinates.\\
The aim of this section is to state the form of spinorial fields, on a space-time manifold with a compactified three-dimensional sphere.\\ The relic un-broken symmetry group in the full tangent space is $SO(1;3)\otimes SO(3)$ and therefore we can build up an eight-component representation in the following way:
\begin{equation}
\Psi_{r}=\chi_{rs}\psi_{s}\qquad r,s=1,2
\end{equation}
$\chi$ being a $SU(2)$ representation and $\psi_{s}$ Dirac spinors.\\
Hence, we assume to factorize the dependence on four-dimensional and extra-dimensional coordinates (denoted by x and y respectively), such that $\chi=\chi(y)\quad\psi=\psi(x)$.\\
A physical spinor field should arise as the solution of the corresponding Dirac equation, which takes the following form in such a KK manifold    
\begin{eqnarray}
\chi\gamma^{(\mu)}(e_{(\mu)}^{\mu}\partial_{\mu}-\Gamma_{(\mu)})\psi+\gamma^{(\mu)}e_{(\mu)}^{m}(\partial_{m}\chi)\psi+\gamma^{(m)}[(e_{(m)}^{m}\partial_{m}-\Gamma_{(m)})\chi]\psi=0\label{eqdir}
\end{eqnarray}
where Greek and Latin letters indicate four-dimensional and extra-dimensional components, respectively, while indices in parenthesis are n-bein ones. For extra-dimensional $\gamma$ matrices we have $\gamma^{(m)}=\gamma_{5}\sigma_{(m)}$, $\sigma_{(m)}$ being Pauli matrices.\\
In equation (\ref{eqdir}) we also assume a non-Riemannian space-time manifold, by taking as spinorial connections just those adapted to the four-dimensional and to the extra-dimensional space, respectively i.e.
\begin{equation}
\left\{\begin{array}{c}\Gamma_{(\mu)}=^4{}\!\Gamma_{(\mu)}\qquad\qquad\\
\Gamma_{(m)}=^3{}\!\Gamma_{(m)}=\frac{i}{2}\sigma_{(m)}\end{array}\right.;
\end{equation}
such a choice is a natural consequence of breaking the Lorentz symmetry into the direct product of two different group, $SO(1;3)$ and $SO(3)$.\\
Now, the effective four-dimensional theory for Dirac spinors is obtained after the computation of $\chi$ terms.\\
To avoid mass terms of the compactification order for $\psi$, we should require $\chi$ to be a solution of the massless Dirac equation on the 3-sphere. However, no exact solution of the massless Dirac equation on the three sphere (and in general on a compact manifold \cite{13}) exists.
However, phenomenologically, we do not need so much; in fact, when we consider the reduction of a multidimensional theory, the un-observability of extra-dimensions has to be taken into account. The dynamics on $S^{3}$ being undetectable, we have to integrate on such a space to perform the correct splitting \cite{7}, i.e. we must take an average procedure on the extra-coordinates dependence, which, for the extra-space homogeneity, corresponds to a unit weight.\\
Thus, we look for a solution of the following averaged Dirac equation
\begin{equation}
\int_{S^{3}} d^{3}y\sqrt{\gamma}\gamma^{(m)}(e_{(m)}^{m}\partial_{m}-\frac{i}{2}\sigma_{(m)})\chi=0.\label{dirint}
\end{equation}
In previous works \cite{6}\cite{7}\cite{8}, we considered the following form for $\chi$
\begin{equation}
\chi_{r}=\frac{1}{\sqrt{V}}e^{-\frac{i}{2}\sigma_{(p)rs}\lambda^{(p)}_{(q)}\Theta^{(q)}(y^{m})}\label{sp}
\end{equation}
with $V$ the volume of $S^{3}$ and $\lambda$ a constant matrix satisfying 
\begin{equation}
(\lambda^{-1})^{(p)}_{(q)}=\frac{1}{V}\int_{S^{3}}
\sqrt{-\gamma}e^{m}_{(q)}\partial_{m}\Theta^{(p)}d^{3}y.
\end{equation}
Because of the arbitrariness of the $\Theta$ functions, we implicitly assumed the commutativity between $\chi$ and its partial derivatives. However, this request implies $\lambda^{-1}$ to have a vanishing determinant. In fact, if $\chi$ and $\partial_m\chi$ commute, it can be shown by some algebra that $\Theta^{(q)}=f(y)c^{(q)}$, $c^{(q)}$'s being constants, and therefore we could get 
\begin{equation}
(\lambda^{-1})^{(p)}_{(q)}=c^{(p)}\frac{1}{V}\int_{S^{3}}\sqrt{-\gamma}e^{m}_{(q)}\partial_{m}f d^{3}y=c^{(p)}d_{(q)}
\end{equation}
whose determinant vanishes.\\
Here we allow $c^{(p)}$ to have a dependence on y-coordinates, instead, but we control the commutativity by virtue of an order parameter, i.e. we can require the validity of the picture (\ref{sp}) to an arbitrary degree of approximation. Hence, we write 
\begin{equation}
\label{theta}\Theta^{(p)}=\frac{1}{\beta}c^{(p)}e^{-\beta\eta}\qquad\eta>0
\end{equation}
and, without loss of generality, we require that $\partial_{m}c^{(p)}\sim c^{(p)}\partial_{(m)}\eta$, so that
\begin{equation}
\partial_{m}\chi=\frac{i}{2}\beta\bigg[\sigma_{(p)}\chi\lambda^{(p)}_{(q)}c^{(q)}\partial_{m}\eta+O(\beta^{-1})\bigg].
\end{equation}
By substituting the last expression into equation (\ref{dirint}), and by expanding $\chi$ in a series of the parameter $\beta$, we find 
\begin{equation}
\int_{S^{3}} d^{3}y\sqrt{\gamma}\gamma^{(m)}e_{(m)}^{m}\partial_{m}\chi=\frac{i}{2}\sigma_{(m)}\chi+O(\beta^{-1})
\end{equation}
thus, the bigger the parameter $\beta$ is, the better the spinor approximates the solution of the massless Dirac equation. We stress that, in the above mentioned previous works, we had to neglect $\chi$ to recover the spinor (\ref{sp}) as a solution of the Dirac equation.\\  
Moreover, since $\beta$ determines the size of the phase, it is related to the extra-dimensional momentum; but the amount of energy needed to excite extra-dimensional modes is of the order of the compactification scale, so that, in the low energy limit, a weak dependence on extra-coordinates (i.e. a big value for $\beta$) is expected.\\ 
Finally, it is clear that we are performing a low energy effective theory, since the spinor is no longer a good approximate solution as we approach the compactification energy scales. Therefore, we do not need to discuss the renormalizability of our model.

\section{Geometrization of the $SU(2)$ gauge connection}
In a Kaluza-Klein scheme we can geometrize boson degrees of freedom, but, unless we introduce Supersymmetries, fermions have to be treated as matter fields. So, we can state that the geometrization of interactions is performed only after recovering gauge connections starting from free spinors.\\
Let us assume KK hypothesis for the space-time manifold $V^{4}\otimes S^{3}$, i.e.
\begin{equation}
\label{m1}\left\{\begin{array}{c} g_{\mu\nu}=\eta_{(\mu)(\nu)}e_{\mu}^{(\mu)}e_{\nu}^{(\nu)}\\
e_{\mu}^{(m)}=\alpha W^{(m)}_{\mu}\\
e_{m}^{(\mu)}=0\\\gamma_{mn}=\eta_{(m)(n)}e_{m}^{(m)}e_{n}^{(n)}\end{array}\right.,
\end{equation}
$g_{\mu\nu}$ and $\gamma_{mn}$ being the four-dimensional and the extra-dimensional metric, respectively, with $\alpha$ giving the length of the extra-space, while $W^{(m)}_{\mu}$ are gauge bosons and  $e^{m}_{(m)}$ 3-bein vectors on $S^{3}$ which are Killing vectors of the 3-sphere (for their explicit form see \cite{9}). Thus, the set $\{e^{m}_{(m)}\}$ satisfies the algebra of the $SU(2)$ group
\begin{equation}
e^{n}_{(n)}\partial_{n}
e_{(m)}^{m}-e^{n}_{(m)}\partial_{n}
e_{(n)}^{m}=\frac{1}{\alpha}C^{(p)}_{(n)(m)}e^{m}_{(p)},
\end{equation}
with $C^{(p)}_{(n)(m)}$ structure constants.\\
Let us consider the free Dirac action
\begin{equation}   
S=\frac{i\hbar c}{2}
\int_{V^{4}\otimes S^{3}}[D_{(A)}\bar{\Psi}\gamma^{(A)}\Psi-\bar{\Psi}\gamma^{(A)}D_{(A)}\Psi]\sqrt{g}\sqrt{\gamma}d^{4}xd^{3}y
\end{equation}
together with the form of the spinor found in the previous section; after the dimensional reduction, we obtain at the lowest order in $\beta^{-1}$ the right gauge coupling, i.e.
\begin{eqnarray*}
S=\frac{i\hbar c}{2}\int
\bar{\psi}\bigg[\gamma^{(\mu)}\bigg(D_{(\mu)}-\frac{i}{2\alpha}e^{\mu}_{(\mu)}W_{\mu}^{(m)}\sigma_{(m)}\bigg)
\psi-c.c.+O(\beta^{-1})\bigg]\sqrt{-g}d^{4}x.
\end{eqnarray*}
This way, we showed that it is possible to geometrize in the low-energy limit a SU(2) gauge theory.\\
However, a clear indication of additional higher order terms in $\beta$ (and not removable by the manifold symmetries) appears; the request to disregard these terms will give a lower bound for the parameter $\beta$ in the application to the Standard Model.\\
Now, we can say that the effect of such terms is to break the gauge invariance by adding an interaction of the following type
\begin{equation}
\ldots+\frac{i}{2\beta}W_{\mu}^{(m)}M^{(n)}_{(m)}\bar{\psi}\sigma_{(n)}\psi+\ldots\label{adter}
\end{equation}
$M^{(n)}_{(m)}$ being  
\begin{equation}
M^{(n)}_{(m)}\sigma_{(n)}=\frac{1}{V}\int_{S^{3}}d^{3}y\chi^{\dag}\int_0^1 ds\chi^s\sigma_{(r)}\lambda^{(r)}_{(s)}e^{-\eta\beta}
e^{m}_{(m)}\partial_m c^{(s)}\chi^{1-s}  
\end{equation}
a constant matrix with no obvious properties.

\section{Geometrization of the Electro-Weak Model}
We now apply results of previous sections to the Electro-Weak Model.\\ 
First of all, let us consider the space-time manifold $V^{4}\otimes S^{1}\otimes S^{3}$, where the geometrization of a $SU(2)\otimes U(1)$ gauge theory can be performed in a KK background \cite{8}. In this sense, we assume the well-known form for the metric tensor, where gauge bosons $W^{(m)}_{\mu}$ and $B_{\mu}$ arise as off-diagonal components
\begin{equation}
j_{AB}=\left(\begin{array}{c|c|c}g_{\mu\nu}+B_{\mu}B_{\nu}+\delta_{(m)(n)}W^{(m)}_{\mu}W^{(n)}_{\nu}
& \alpha'B_{\mu} & \alpha e_{m(m)}W^{(m)}_{\mu} \\
\hline\\
\alpha'B_{\nu} & 1 & 0 \\\hline\\
\alpha e_{n(n)}W^{(n)}_{\nu}
& 0 & \gamma_{mn}\end{array}\right).
\end{equation}
with $\alpha'$ giving the length of the $S^{1}$ space.\\
The main point is that by the dimensional reduction of the Einstein-Hilbert action
\begin{equation}
\label{k1}S=-\frac{c^{4}}{16\pi G_{(n)}}\int_{V^{4}\otimes S^{1}\otimes S^{3}}{}^n\!R\sqrt{-j}
d^{4}xdy^{0}d^{3}y
\end{equation}
the Yang-Mills Lagrangian density for gauge bosons $B_{\mu}$ and $W^{(m)}_{\mu}$ comes out
\begin{eqnarray*}
S=-\frac{c^{3}}{16\pi G}\int_{V^{4}}
\sqrt{-g}\bigg[R-\frac{1}{4}B_{\mu\nu}B_{\rho\sigma}g^{\mu\rho}g^{\nu\sigma}-\frac{1}{4}\delta_{(n)(m)}F^{(n)}_{\mu\nu}F^{(m)}_{\rho\sigma}g^{\mu\rho}g^{\nu\sigma}+R_{1+3}\bigg],
\end{eqnarray*}
$R_{1+3}$ being the curvature of the extra-dimensional space. By the Standard Weinberg rotation, Z and photon fields arise.\\
Once spinors have been introduced, we account for isospin degrees of freedom by a dependence on $S^{3}$ coordinates of the form (\ref{sp}). Instead, as usual in KK theories, the geometrization of a $U(1)$ gauge connection is easily obtained by a phase dependence on the $S^{1}$ variable.\\
Finally, we are able to reproduce any quark generation and fermion family of the Standard Model by the following multidimensional spinors 
\begin{equation}   
\Psi_{Lr}(x;y;\theta)=e^{in_{s}\theta}\chi_{rs}(y)\psi_{Ls}(x)\qquad
\Psi_{Rr}(x;y;\theta)=e^{in_{r}\theta}\psi_{Rr}(x);\qquad \theta=\frac{y^{0}}{\alpha'}\qquad Y_{r}=\frac{n_{r}}{6}
\label{LRfields}
\end{equation}
$Y_{r}$ being the hypercharge of the four-dimensional spinor $\psi_{r}$, while $\psi_{R}$ and $\psi_{L}$ stand for the two four-dimensional chirality ($\gamma_{5}$) eigenstates. In the present phenomenological approach, the chirality problem is overcome because we can deal directly with four-dimensional fields; such 4-spinors are the relic of multidimensional states characterized by a different dependence on extra-dimensions (accounting for the different isospin numbers). \\ 
Now, for left-handed fields we have to identify $\psi_{L}$ with isospin doublets. Therefore, it is a natural choice to think of the two components of $\psi_{R}$ as right-handed partners of each doublet.\\
Since we developed our model in an eight-dimensional space-time, we deal with multidimensional spinors having 16 components. We suggest to recast these components, so that any spinor contains a quark generation and a fermion family, i.e.
\begin{equation} 
\Psi_{L}=\frac{1}{\sqrt{V\alpha'}}\left(\begin{array}{c} \chi\left(\begin{array}{c} e^{in_{uL}\theta}u_{L}\\e^{in_{dL}\theta}d_{L}\end{array}\right)\\\chi\left(\begin{array}{c} e^{in_{\nu L}\theta}\nu_{eL}\\e^{in_{eL}\theta}e_{L}\end{array}\right)\end{array}\right)\qquad\Psi_{R}=\frac{1}{\sqrt{V\alpha'}}\left(\begin{array}{c} \left(\begin{array}{c} e^{in_{uR}\theta}u_{R}\\e^{in_{dR}\theta}d_{R}\end{array}\right)\\\left(\begin{array}{c} e^{in_{\nu R}\theta}\nu_{eR}\\e^{in_{eR}\theta}e_{R}\end{array}\right)\end{array}\right). 
\end{equation}
This way, from 6 multidimensional spinors, one for each family or generation and one for each chirality eigenstate, we are able to reproduce all Standard Model particles. Furthermore, in doing that, we give an explanation for the equal number of quark generations and fermion families.\\

\paragraph{Chirality feature of the model}
The assignment of the above dependence for left-handed and right-handed fields could look {\it ad-hoc} to reproduce quantum numbers of Standard Model particles, but an explanation for it comes from the massless character of the involved fields. In fact, since masses are going to be generated by an Higgs-like mechanism, then the evolution of right-handed and left-handed fields is totally uncorrelated in a 4-dimensional background, before the spontaneous symmetry breaking to take place. This conclusion is modified by terms of the Dirac equation due to the presence of the extra-space. This can be recognized from their explicit form, which reads as follows (for the representation of Dirac matrices see \cite{8})
\begin{equation}
\gamma^{(m)}D_{(m)}\Psi+\gamma^{(7)}D_{(7)}\Psi=\left(\begin{array}{c} \sigma^{(m)}D_{(m)}\left(\begin{array}{c} \gamma_5\Psi_1\\ \gamma_5\Psi_2\end{array}\right)\\\sigma^{(m)}D_{(m)}\left(\begin{array}{c} \gamma_5\Psi_3\\\gamma_5\Psi_4\end{array}\right)\end{array}\right)+\left(\begin{array}{cc} 0 & I\\ I & 0 \end{array}\right)\left(\begin{array}{c} D_{(7)}\left(\begin{array}{c} \Psi_1\\ \Psi_2\end{array}\right)\\D_{(7)}\left(\begin{array}{c} \Psi_3\\\Psi_4\end{array}\right)\end{array}\right)
\end{equation}
and clearly indicates that starting from left-handed fields, mixing terms with right-handed ones are present in the equation giving the dynamics.\\
Despite the proper definition of chirality in a multi-dimensional scenario, an independent dynamics for right-handed and left-handed fields can be reproduced by treating 4-dimensional chirality eigenstates like fundamental fields. In this respect, to assume a different dependence on extra-coordinates is then required in order to restore the proper 4-dimensional phenomenology, {\it i.e.} the chirality of the SU(2) gauge interaction. Furthermore, it also avoids the emergence of a correlation between the two 4-chirality eigenstates and, at the same time, the appearance of huge Kaluza-Klein mass terms. In fact, in order to reproduce proper quantum numbers, one ends up with terms of the form 
\begin{equation}  
\bar{\Psi}\gamma^{(7)}\Psi=\left(\begin{array}{cccc} (\bar{\Psi}_1 & \bar{\Psi}_2) & (\bar{\Psi}_3 & \bar{\Psi}_4) \end{array}\right)\left(\begin{array}{cc} 0 & I\\ I & 0 \end{array}\right) \left(\begin{array}{c}\left(\begin{array}{c}\Psi_1\\ \Psi_2 \end{array}\right) \\ \left(\begin{array}{c}\Psi_3\\ \Psi_4\end{array}\right)\end{array}\right)
\end{equation}
which vanishes, at least classically, as far as the field $\Psi$ is a chirality eigenstate.\\ 
All these speculations clearly indicate that a deep connection exists between the avoidance of Kaluza-Klein mass terms and the chirality issue. 

\paragraph{}
From the results of previous sections, one can easily recognize that starting from the Dirac Lagrangian density for each spinor
\begin{equation}  
S_{\Psi}=-\frac{i\hbar c}{2}\int_{V^{4}\otimes S^{1}\otimes S^{3}}
[\bar{\Psi}_{L}\gamma^{(A)}D_{(A)}\Psi_{L}+\bar{\Psi}_{R}\gamma^{(A)}D_{(A)}\Psi_{R}+c.c.]\sqrt{-\gamma g}d^{4}xd\theta d^{3}y
\end{equation}
we obtain, by dimensional reduction, an effective action containing the gauge interaction, plus additional terms of $\beta^{-1}$  order, i.e.
\begin{eqnarray}
S_{\Psi}=-\frac{i\hbar c}{2}\int_{V^{4}}
[\bar{\psi}_{L}\gamma^{(\mu)}(D_{(\mu)}+\frac{i}{2\alpha}e^{\mu}_{(\mu)}W_{\mu}^{(m)}\sigma_{(m)}+\frac{i}{2\alpha'}y_{L}e^{\mu}_{(\mu)}B_{\mu})\psi_{L}+
\bar{\psi}_{R1}\gamma^{(\mu)}(D_{(\mu)}+\nonumber\\+\frac{i}{2\alpha'}y_{R1}e^{\mu}_{(\mu)}B_{\mu})\psi_{R1}+\bar{\psi}_{R2}\gamma^{(\mu)}(D_{(\mu)}+\frac{i}{2\alpha'}y_{R2}e^{\mu}_{(\mu)}B_{\mu})\psi_{R2}+c.c.+O(\beta^{-1})]\sqrt{-\gamma g}d^{4}x
\end{eqnarray}
Hence, we have just performed the geometrization of the Electro-Weak interaction for spinors.\\
Then, we introduce coupling constants by redefining gauge bosons
\begin{equation}
B_{\mu}\Rightarrow kg'B_{\mu}\qquad
W^{(i)}_{\mu}\Rightarrow kg W^{(i)}_{\mu}\quad (i=1,2,3)
\end{equation}
and, by imposing the Lagrangian density to coincide with the Electro-Weak model one, the following relations between coupling constants and extra-dimensions lengths arise
\begin{equation}
\alpha^{2}=16\pi G\bigg(\frac{\hbar}{gc}\bigg)^{2}=0.18\times 10^{-31}cm\qquad
\alpha'^{2}=16\pi G\bigg(\frac{\hbar}{g'c}\bigg)^{2}=0.33\times10^{-31}cm.
\end{equation}
Because of these estimates, less than two order of magnitude greater than Planck length, we expect we'll be able to explain the stabilization of the extra-space, in a quantum gravity framework (for a classical mechanism of stabilization see \cite{Sal}).

\section{Non standard couplings}
In section 2 we have found a solution of the Dirac equation on the three-sphere at the lowest order in $\beta^{-1}$ and we have achieved the geometrization of the Electro-Weak interaction. At higher orders, there is no clear indication that the geometrization of connections can be performed by virtue of the Dirac equation solution, therefore we expect some deviations (of order $\beta^{-1}$) to occur on the $SU(2)$ component of the Electro-Weak model.\\
In particular, these additional terms are of the form ($\ref{adter}$) and they introduce new vertexes, which break gauge symmetries. We can get an estimate of the parameter $\beta$ by considering that these vertexes will induce modification on cross sections. These modifications are of the $\beta^{-1}$ order for Standard Model processes, because of interferences, while probabilities for forbidden decays will be $O(\beta^{-2})$. This fact enables us to estimate a lower bound for $\beta$ from current limits on forbidden decays, since contributions to precision tests, like corrections to partial widths of W and Z, will result to be far below the experimental uncertainty.\\
Let us consider the partial width for the decay of a neutron in a proton plus a couple of neutrino-antineutrino, which actually has the following experimental limit \cite{10}
\begin{equation}
\Gamma(n\rightarrow p+\nu_{e}+\bar{\nu}_{e})/\Gamma_{tot}<8*10^{-27}.  
\end{equation}
Since such a decay breaks the electric charge conservation, its Feynman diagrams must contain at least one vertex responsible for the violation, i.e. of the form ($\ref{adter}$).\\

\begin{figure}[!htb]
\begin{center}
\begin{tabular}{cccccccccccccccc}    
Fig. 1 
   
\begin{fmffile}{one} 	
  \fmfframe(1,7)(1,7) 	{
   \begin{fmfgraph*}(110,62) 
    \fmfleft{i1,i2}	
    \fmfright{o1,o2}    
    \fmflabel{$d$}{i1} 
    \fmflabel{$u$}{i2} 
    \fmflabel{$\bar{\nu}$}{o1} 
    \fmflabel{$\nu$}{o2} 
    \fmf{fermion}{i1,v1,i2} 
    \fmf{fermion}{o1,v2,o2} 
    \fmf{photon,label=$Z$}{v1,v2} 
   \end{fmfgraph*}
 } 
\end{fmffile}
\end{tabular}

\end{center}

\end{figure}

For instance, a possible Feynman diagram is in figure 1, where we have a vertex with Z and a couple of quarks u-d; by substituting into $W^{3}_{\mu}$ the photon and the Z fields, we obtain, among the others, a term describing this coupling in the expression ($\ref{adter}$) , i.e. 
\begin{equation}
\ldots+\frac{i}{2\beta}\cos\theta_{W}Z_{\mu}\bar{\psi}(M^{(1)}_{(3)}\sigma_{(1)}+M^{(2)}_{(3)}\sigma_{(2)})\psi+\ldots
\end{equation}
Therefore, the amplitude for the process will be of order of $\beta^{-1}$, the cross section being O($\beta^{-2}$) and, this way, we get
\begin{equation}
\frac{1}{\beta^2}\leq 10^{-27}\Rightarrow\beta\geq10^{14}.
\end{equation}
With such a big value for $\beta$, corrections to the cross sections of the allowed Standard Model processes, coming out from interferences with new vertexes, are suppressed far-below the present experimental uncertainty for recent electroweak precision tests, too. For instance, the partial widths of Z and W receive corrections of the $\beta^{-1}\leq 10^{-14}$, which is far below the experimental uncertainty (which is at most 0$(10^{-4})$).

\section{The Higgs mechanism}
In order to give masses to particles, we have to introduce a field, whose dimensional reduction gives the Higgs boson.\\
Let us consider the following form for such a multi-scalar field
\begin{equation}
\label{Phi}\Phi=\left(\begin{array}{c}\Phi_{1}
\\\Phi_{2}
\end{array}\right)=\frac{1}{\sqrt{V\alpha'}}\chi\left(\begin{array}{c}e^{-3i\theta}\phi_{1}
\\ e^{3i\theta}\phi_{2}
\end{array}\right)
\end{equation} 
from which it can be recognized $\phi_{1}$ and $\phi_2$ hypercharges are $-\frac{1}{2}$ and $\frac{1}{2}$, respectively.\\
Now, starting from a Higgs-like Lagrangian density, i.e.
\begin{equation}
\label{lhig}\Lambda_{\Phi}=\frac{1}{2}\eta^{(A)(B)}\partial_{(A)}\Phi^{\dag}\partial_{(B)}\Phi
-\mu^{2}\Phi^{\dag}\Phi-\lambda(\Phi^{\dag}\Phi)^{2},
\end{equation}
we get (at the lowest order in $\beta^{-1}$) the 4-dimensional Higgs Lagrangian density
\begin{eqnarray*}
S_{\Phi}=\frac{1}{c}\int d^{4}x \sqrt{-g}
\bigg[\frac{1}{2}g^{\mu\nu}(D_{\mu}\phi)^{\dag}D_{\nu}\phi-
\mu^{2}\phi^{\dag}\phi-\lambda(\phi^{\dag}\phi)^{2}+\frac{1}{2}
G\phi^{\dag}\phi+O(\beta^{-1})\bigg].
\end{eqnarray*}
as soon as we impose the following condition
\begin{equation}
\lambda^{(p)}_{(q)}\lambda^{(r)}_{(s)}\frac{1}{V}\int_{S^{3}}\sqrt{\gamma} c^{(q)}c^{(s)}\xi^{m}_{(m)}\partial_{m}\eta\xi^{n}_{(n)}\partial_{n}\eta e^{-2\beta\eta}d^{3}y=\delta^{(p)}_{(m)}\delta^{(r)}_{(n)}.
\end{equation}
We emphasize that there is an additional mass term with a coefficient $G$ which, in geometric units, is of the compactification scale order, thus it can explain the fine-tuning requested to stabilize the Higgs mass \cite{sus}; in fact, it is obvious that, in such a framework, a natural cut-off is given by the extra-dimension length, since, at such scales, the present form of the spinor (\ref{sp}) can no longer be justified. The cancellation between the radiative corrections to Higgs mass and the massive term, here appearing, takes place in view of the opposite sign of these contributions and the expected comparable amplitudes.\\
On this level, we are now ready to give masses to gauge bosons by the usual spontaneous symmetry breaking mechanism. However, the following $\beta^{-1}$ corrections have to be taken into account
\begin{equation}
\ldots+2\beta^{-1}M^{(n)}_{(m)}W^{(m)}_{\mu}W^{(r)}_{\nu}g^{\mu\nu}\phi^\dag\sigma_{(n)}\sigma_{(r)}\phi+2\beta^{-1}N_{(m)(s)}W^{(m)}_{\mu}W^{(s)}_{\nu}g^{\mu\nu}\phi^{\dag}\phi+\ldots
\end{equation}
where
\begin{eqnarray}
N_{(m)(s)}=\frac{1}{V\alpha'}\delta_{(q)(t)}\lambda^{(q)}_{(u)}\lambda^{(t)}_{(v)}\int d^{3}yd\theta \sqrt{\gamma}e^{m}_{(m)}\partial_m c^{(u)}e^{s}_{(s)}\partial_s\eta 
c^{(v)}e^{-2\beta\eta}\chi^{s}.
\end{eqnarray}
Thus, we get a non-diagonal mass matrix for bosons, therefore corrections to their masses and new (and in general gauge-violating) anomalous self-interactions arise.\\
From current limits on the photon mass \cite{10}
\begin{equation}
m_{\gamma}<6*10^{-17}eV\label{lowbou}
\end{equation}
we obtain a lower bound for $\beta$ greater than the one of the previous section
\begin{equation}
\frac{1}{\beta}\sim\frac{m_{\gamma}}{m_{Z}}\Rightarrow\beta>10^{28}.
\end{equation}
This lower bound is much greater than the one coming from precision electroweak tests. For instance, from the measured value of the W mass $m_{W}=(91,1876\pm0.0021)GeV$, we infer that $\beta^{-1}$ corrections are negligible for $\beta>>10^{10}$.\\  
Furthermore, mass terms for spinors by standard Yukawa couplings can be inferred; we can rewrite these terms in the following way
\begin{eqnarray}
g[\bar{\Psi}_{lL}\Phi\Psi_{lR}+c.c.]=g\sum_{r=1}^{4}[(\bar{\Psi}_{1})_{r}\Phi_{1}(\Psi_{1})_{r}+(\bar{\Psi}_{1})_{4+r}\Phi_{2}(\Psi_{2})_{4+r}+\nonumber\\+(\bar{\Psi}_{1})_{8+r}\Phi_{1}(\Psi_{lR})_{8+r}+(\bar{\Psi}_{1})_{12+r}\Phi_{2}(\Psi_{2})_{12+r}+c.c.]\qquad\qquad
\end{eqnarray}
We outline that, since different hypercharges of Higgs components are here allowed, then gauge invariant mass terms for neutrinos can be reproduced this way. In the Standard Model, we have to impose the same hypercharge for the Higgs doublets, in order to get the commutativity between the $SU(2)$ and the $U(1)$ groups \cite{Man}. Here, we do not need this request, since the action of these groups are reproduced by translations on different spaces, so that they commute automatically.\\ 
Moreover, any vacuum expectation value preserves the electric charge conservation, at the lowest order in $\beta^{-1}$. In fact, from a four-dimensional perspective, the generator of hypercharge transformations on Higgs field is the opposite of the one associated with the third component of the isospin. Finally, we are free to take any vacuum expectation value, i.e.
\begin{equation}
\phi=\left(\begin{array}{c} v_{1}+\sigma(x) \\ v_{2} \end{array}\right) \quad v_{1}=v\cos\varphi\quad v_{2}=v\sin\varphi.
\end{equation} 
In order to get the mass spectrum of Standard Model particles, we have to redefine any four-dimensional field by a phase $c_{\psi}$
\begin{equation}
\psi\rightarrow c_{\psi}\psi \qquad c^{*}_{\psi}c_{\psi}=1
\end{equation}   
whose relation with masses (achieved after the spontaneous symmetry breaking) reads
\begin{equation}
m_{\psi}=gv_{1/2}(c^{*}_{\psi R}c_{\psi L}+c^{*}_{\psi L}c_{\psi R})
\end{equation}
where $v_{1}$ and $v_{2}$ applies to the first and to the second isospin components, respectively.\\
We conclude this section by stressing the new issues of our approach, when the spontaneous symmetry breaking is considered:\\
(i) no fine-tuning on the Higgs parameters is required any more in order to explain the radiative contribution; if we take the natural cut-off of our model (two or three order the Planck mass), then the real Higgs mass is expected to have a much lower amplitude in view of the cancellation with the Kaluza-Klein massive term (generated after the dimensional reduction);\\ 
(ii) the greater freedom we have in fixing the Higgs hypercharge and vacuum state allows us to deal with a non-zero neutrino mass, in the framework of a standard Yukawa coupling. 

\section{Conclusions}
In this paper, we have considered a spinor defined on a three dimensional sphere. We have found that a low energy solution of the Dirac equation is suitable to geometrize $SU(2)$ gauge connections. We have expressed the low energy limit by virtue of a parameter $\beta$, which determines the dependence of the spinor on the extra-space (bigger $\beta$ is, less sensitive the spinor is to extra-coordinates). We have found that gauge symmetries are broken, even at low energies, but that terms describing this kind of violations are suppressed by a factor $\beta^{-1}$. Then, we have applied these results to the specific case of the usual Electro-Weak model, i.e. the $U(1)$ hypercharge symmetry is included in the theory. Thus, analyzing the implication of $\beta$-terms on the $SU(2)$ sector, lower bound for $\beta$ from current limits on Standard Model symmetries came out.\\
Moreover, we have also considered the application of the Higgs mechanism in this context. First of all, we have taken an Higgs field having two components with opposite hypercharges. In doing that, we got the new relevant issue of a non-zero neutrino mass by standard Yukawa terms. Thus, $SU(2)$ properties of the Higgs field have been reproduced by virtue of the same dependence on $S^{3}$ coordinates as for spinors: although this assumption have not come out from the field equation (as it was for spinors), nevertheless we have been forced to consider it as preserving the Electro-Weak symmetries invariance and, in this sense, it has worked suitably. In fact, after the dimensional reduction, the Higgs-Yang-Mills Lagrangian density has come out, together with a mass terms of the compactification length order, which can explain the well-known Higgs mass fine-tuning.\\
Gauge-violating $\beta^{-1}$ terms have provided new type of interactions among gauge bosons and corrections to their masses, including a mass term for photons. The bigger limit we got for $\beta$ (\ref{lowbou}) has come just from experimental limits on the photon mass.\\ 
We regard the possibility to preserve the electric charge conservation as subject of future investigations, thus getting a much smaller lower bound for $\beta$. Nevertheless, such a large value of $\beta$ is not surprising since we are far away from energy scales of the compactification length order; i.e. we estimate that the probability to excite extra-dimensional modes is highly suppressed and, as a consequence, a weak dependence on extra-coordinates for matter fields has to take place.\\

\end{document}